\documentclass{article}

\usepackage[preprint, nonatbib]{nips_2018}

\usepackage[utf8]{inputenc} % allow utf-8 input
\usepackage[T1]{fontenc}    % use 8-bit T1 fonts
\usepackage{hyperref}       % hyperlinks
\usepackage{url}            % simple URL typesetting
\usepackage{booktabs}       % professional-quality tables
\usepackage{amsfonts}       % blackboard math symbols
\usepackage{nicefrac}       % compact symbols for 1/2, etc.
\usepackage{microtype}      % microtypography
\usepackage{algorithm}
\usepackage{algpseudocode}
\usepackage{graphicx}
\usepackage{amsmath}
\usepackage{color}
\usepackage{diagbox}

\usepackage{amsthm}

\newcommand{\R}{{\mathbb{R}}}
\newcommand{\N}{{\mathbb{N}}}
\newcommand{\ip}[2]{\langle #1,#2 \rangle}

\newtheorem*{thm*}{Theorem}
\newtheorem*{prop*}{Proposition}

\title{Spike Sorting by Convolutional Dictionary Learning}

\author{
  Andrew H. Song\\
  Department of EECS\\
  MIT\\
  Cambridge, MA 02139 \\
  \texttt{andrew90@mit.edu} \\
  \And
  Francisco Flores\\
  Department of Anesthesia, Critical Care and Pain Medicine\\
  Massachusetts General Hospital and Harvard Medical School\\
  Boston, MA 02114 \\
  \texttt{fjflores@neurostat.mit.edu} \\
  \And
  Demba Ba\\
  School of Engineering and Applied Sciences\\
  Harvard University\\
  Cambridge, MA 02138 \\
  \texttt{demba@seas.harvard.edu} \\
}

\begin{document}

\maketitle

\begin{abstract}
Spike sorting refers to the problem of assigning action potentials observed in extra-cellular recordings of neural activity to the neuron(s) from which they originate. We cast this problem as one of learning a convolutional dictionary from raw multi-electrode waveform data, subject to sparsity constraints. In this context, sparsity refers to the number of neurons that are allowed to spike simultaneously. The convolutional dictionary setting, along with its assumptions (e.g. refractoriness) that are motivated by the spike-sorting problem, let us give theoretical bounds on the sample complexity of spike sorting as a function of the number of underlying neurons, the rate of occurrence of simultaneous spiking, and the firing rate of the neurons. We derive memory/computation-efficient convolutional versions of OMP (cOMP) and KSVD (cKSVD), popular algorithms for sparse coding and dictionary learning respectively. We demonstrate via simulations that an algorithm that alternates between  cOMP and cKSVD can recover the underlying spike waveforms successfully, assuming few neurons spike simultaneously, and is stable in the presence of noise. We also apply the algorithm to extra-cellular recordings from a tetrode in the rat Hippocampus.

\end{abstract}

\section{Introduction}

In experimental neuroscience, electrophysiology using extra-cellular electrodes has been the de-facto method to record neural activity from brain. With the falling costs of storage, a recent trend is the collection and storage of raw extracellular neural activity using large electrode arrays, comprising hundreds to thousands of electrodes~\cite{Rey2015}, for prolonged period of up to hours/days \cite{Dhawale2017}, and at high sampling rates. These developments have enabled the probing of neural dynamics at large spatial and temporal scale, and have shown promising improvements in spike sorting, that is, the association of action potentials from extra-cellular recordings to the neuron(s) from which they originate. 
%As such, there has been growing interest in scalable approaches to processing raw  extra-cellular waveform data.  % Henceforth, we will refer to action potentials and spikes interchangeably.

Reflecting this trend, numerous approaches for spike sorting have been introduced. They fall into two broad categories. The first approach, based on the clustering of features extracted from the detected spike waveforms, has been the mainstream approach for decades \cite{Lewicki}. The features range from simple spike characteristics, such as peak amplitude and width, to more complicated features such as principal components and wavelet coefficients~\cite{Quiroga}. Recently introduced algorithms such as Mountainsort~\cite{Chung2017} and Spyking Circus \cite{spyking} employ a similar approach, with sophisticated metrics and checks to prevent spurious events from affecting the clustering.

The second approach, fairly recent compared to the first, is focused on learning (or discovering) the signature/template waveform from each of the neurons sensed by a given electrode array, and on using these to find the location of the action potentials that best match the learned signature waveforms. Typically, the templates are iteratively learned, for instance by computing a running average of the waveforms classified as coming from the same neuron \cite{kilosort}. The Matching Pursuit algorithm~\cite{Mallat1993} has been the popular method for matching the templates and identifying the neuron(s) associated with extra-cellular action potentials~\cite{spyking,kilosort,  YASS}.  

More recently, the second approach has been strengthened by drawing from the signal processing literature, particularly the sparse approximation literature \cite{TROP2007}. A simple generative model for the observations from a single extra-cellular electrode is the sum of the convolution of each spike waveform with the marked point-process consisting of the spike times from a given neuron and the associated amplitude \cite{Sahani1997}. In the context of this generative model,~\cite{ekanadham2011blind, Simoncelli} cast the problem of learning both the spike waveforms and the spike-time/amplitude pairs as a bi-convex optimization problem. This leads to an algorithm, Continuous Basis Pursuit (CBP), that alternates between a sparse-approximation step to identify spike-times/amplitudes given approximate spike waveforms, and a step that updates the spike waveforms given improved spike times/amplitudes from the sparse approximation step. The CBP approach is computationally demanding as it requires the solution to large-scale convex optimization problems.
%Fast approaches (Kilosort) are not grounded in sparse coding/dictionary learning theory and therefore cannot provide theoretical bounds in sense as above.
%Overlapping spike problems - The convolutional framework gracefully deals with the spike overlapping problem, as it is able to delineate different spike waveforms contributing to the overlapped segment.
%
%Moreover, the previous works fail to address any theoretical guarantees on the required amount of data for successful recovery of the spike waveforms and its sparse codes, arguably an important consideration for spike sorting applications. 

Here, we cast spike sorting as a convolutional dictionary learning problem and propose an efficient iterative alternating-minimization algorithm for its solution. At each iteration, the algorithm alternates between convolutional sparse coding via convolutional orthogonal matching pursuit (cOMP) and convolutional dictionary learning via convolutional K-SVD (cKSVD). This follows the general philosophy of the second general approach to spike sorting described above, with the connection to dictionary learning and sparse approximation made explicit. The innovations from our approach are twofold. Firstly, unlike in~\cite{ekanadham2011blind, Simoncelli} and \cite{spyking,kilosort, YASS}, cOMP and cKSVD leverage the convolutional form of the linear operators to perform highly memory and computation efficient operations, which in their naive form would be very slow applied to high-sampling-rate recordings. We use simulated and real data, for which ground-truth intracellular data are available \cite{Henze2000}, to show that the proposed approach is able to learn accurate spike waveforms as well as significantly reduce misclassification errors. Secondly, framing spike-sorting as convolutional dictionary allows us to apply results from dictionary learning theory that were developed only recently \cite{altmin}. Specifically, under some regular assumptions that we argue are reasonable in the spike-sorting setting, we give a theoretical bound for the required number of samples (or recording length) to reliably estimate the spike waveforms of a group neurons. 

\section{Spike Sorting as Convolutional Dictionary Learning}
\label{gen_inst}

\subsection{Generative model and assumptions}

Let $n = 1,\cdots,N \in \N^{+}$ be a discrete-time index and $y[n]$ denote a discrete-time signal that represents the voltage from an extracellular electrode recording neural activity. The electrode is able to reliably capture the activity from $C$ neighboring neurons, each with spike-waveform template $h_c[n]$, $c=1,\cdots,C$. As is standard in the spike-sorting literature, we assume that all the templates have equal length $l$. Letting $h_c = (h_c[0],\cdots,h_c[l-1])^{\text{T}}$, we assume without loss of generality, that $\lVert h_c\rVert_2=1,\forall c$. A simple model for $y[n]$ is that it consist of a linear combination of the time-shifted waveform templates, perturbed by additive white noise $\varepsilon[n]$. Mathematically, we can express this model in terms of the convolution between the templates and code vectors $\{x_{c}[n]\}_{c=1}^{C}$
%Let $c$ be the index of a neuron (total of $C$ neurons) each with its waveform template, $h_c$, each of length $l_c$. Without loss of generality, we assume $\lVert h_c\rVert_2=1,\forall c$. Also, let $N_c$ denote the number of times the neuron $C$ spikes within the given data. Our generative model for the voltage trace, $y$, is a linear combination of the time-shifted waveform templates, perturbed by some white noise, $\varepsilon_n$. The linear combination of the time shifts can be expressed as the convolution between the templates and code vectors $\{x_{c}\}_{c=1}^{C}$. The generative model can now be written as
\begin{equation}
y[n] = \sum_{c=1}^C x_{c}[n] \ast h_c[n]+\varepsilon[n],
	\label{eq:dtmodel}
\end{equation}
\noindent where $x_{c}[n] = \sum_{i=1}^{N_c} x_{c,i}\delta[n-n_{c,i}], n=0,\cdots,N-l+1$, and we let $x_c = (x_c[0],\cdots,x_c[N-l+1])^{\text{T}}$.
In practice, the signal $y[n]$ is divided into $J$ non-overlapping windows, each of length $W$ such that $N=WJ$, and $l<<W<<N$. For notational convenience, denote the windowed data by the matrix $Y\in \mathbb{R}^{W \times J}$ whose $j^{\text{th}}$ column $Y_j = (y[(j-1)W],\cdots,y[jW-1])^{\text{T}} \in \R^W$, $j=1,\cdots,J$. %consists of the elements in the $j^{\text{th}}$ window of length $W$ from $y[n]$. 
Further let $x_{c,j}=(x_{c}[(j-1)(W-l+1)],\cdots,x_{c}[j(W-l+1)-1])^\text{T}\in\mathbb{R}^{W-l+1}$ be the code vector for neuron $c$ in window $j$,  $j=1,\cdots,J$. For simultaneous recordings from $M$ electrodes, we partition each electrode into windows in a similar fashion and stack the resulting matrices to obtain $Y\in \mathbb{R}^{W \times MJ}$.

Given $Y$, the goal is to estimate $\{h_c\}_{c=1}^C$ and $\{x_{c,j}\}_{c,j=1}^{C,J}$ that minimize an objective of choice, typically the error in reconstructing $Y$ using the code vectors. Without additional constraints, it is well-known that this is an ill-posed problem, i.e. there does not exist a unique solution. In the context of spike-sorting, nevertheless, we can leverage the following biophysical properties of neurons: 1) Refractoriness prevents the same neuron from spiking again within a certain period ($\approx 1$ ms) and 2) the firing rate for a typical neuron is not high (except for extreme bursting periods). Mathematically, this implies that pairs of elements from $x_{c,j}$ that are close in position cannot both be nonzero and that the total number of nonzero elements of $x_{c,j}$, denoted as $N_{c,j}=\vert \text{supp}(x_{c,j})\vert$, should be small. In other words, $\{x_{c,j}\}_{c,j=1}^{C,J}$ are sparse vectors.

This naturally leads us to incorporate sparsity as a constraint on $x_{c,j}$ to restore well-posedness. Expressed in terms of the $\ell_0$ quasi-norm, $N_{c,j} = \lVert x_{c,j}\rVert_0$. The resulting constrained optimization problem is%We now aim to solve the following optimization problem
\begin{equation}\label{eq:objective}
\min_{\{h_c\}_{c=1}^C,\{x_{c,j}\}_{c,j=1}^{C,J}} \sum_{j=1}^J\Big\lVert Y_j-\sum_{c=1}^C x_{c,j}\ast h_c\Big\rVert_2^2\text{ s.t. } \sum_{c=1}^C\lVert x_{c,j}\rVert_0\leq \beta
\end{equation}  
%The sparse code vector $x_{c,j}$ can be thought of as a collection of $N_{c,j}$ delta functions, the height of which determines the strength of a spike. 
Note that the sparsity constraint in Eq.~\ref{eq:objective} alone does not enforce refractoriness. In practice, we found that enforcing refractoriness is not required explicitly, as it is a salient feature of the data. %prevent the consecutive elements of $x_{c,j}$ from being nonzero, thereby potentially failing to capture the refractoriness. To incorporate this, additional constraints, such as sparsity constraints on the first difference of $x_{c,j}$ has to be imposed, which is not considered here. On a added note, we have observed that the structure of the signal prevents this from happening.

\subsection{Generative model with convolutional dictionary formulation}
In what follows, it will be useful to express Eq.~\ref{eq:objective} in terms of the convolutional dictionary $H$ generated by the templates $\{h_c\}_{c=1}^C$, as follows, where $\lVert \cdot \rVert_F$ denotes the Frobenius norm,
\begin{equation}\label{eq:l0objective}
\min_{H,\{X_j\}_{j=1}^J} \sum_{j=1}^J\Big\lVert Y_j-HX_j\Big\rVert_2^2\text{ s.t. } \lVert X_j\rVert_0\leq \beta \Leftrightarrow \min_{H,X} \Big\lVert Y-HX\Big\rVert_\text{F}^2\text{ s.t. } \lVert X_j\rVert_0\leq \beta
\end{equation}  
$H\in \mathbb{R}^{W\times (\sum_{c=1}^C (W-l+1))}$ is a block-Toeplitz matrix with $C$ blocks $H_1,\cdots,H_C$. $H_c$ is the matrix whose columns consist of all possible timeshifts of $h_c$, zero-padded to have equal length: 
\begin{equation}
H=\begin{bmatrix}
H_1\Big\vert \cdots\Big\vert H_C
\end{bmatrix}
\text{ where }
H_c=\begin{bmatrix}
h_c & 0 & \cdots & 0\\
0 & h_c & \cdots & 0\\
\vdots & 0 & \vdots & \vdots\\
0 & 0 & \cdots & h_c\\
\end{bmatrix}\in \mathbb{R}^{W\times (W-l+1)}
\end{equation}
For each window $j=1,\cdots,J$, the convolutional sparse code $X_j = (x_{1,j}^{\text{T}},\cdots,x_{C,j}^{\text{T}})^\text{T}\in \mathbb{R}^{\sum_{c=1}^C(W-l+1)}$ is a concatenation of the code vectors $\{x_{c,j}\}_{c=1}^C$ from all neurons. Expressing  the convolution operation as a matrix multiplication allows us to seamlessly perform linear algebraic operations such as least-squares. Finally, let $X = \begin{bmatrix} X_1|\cdots|X_J \end{bmatrix} \in \mathbb{R}^{(\sum_{c=1}^C(W-l+1))\times J}$ be the matrix of code vectors from all neurons and all windows and $X^c=[x_{c,1},\cdots,x_{c,J}]\in \mathbb{R}^{(W-l+1)\times J}$ the $c^{th}$ block row of $X$ corresponding to the code vectors from neuron $c$ across all windows.

\subsection{Alternating Minimization}
The objective in Eq.~\ref{eq:l0objective} is nonconvex, due to the simultaneous optimization over $H, X$ and the non-convex $\ell_0$ constraint. A popular approach has been to alternatively minimize the objective over one of the variables while the other is fixed. This process is repeated until a convergence criterion is reached. Let $H^{(t)}$ and $X^{(t)}$ denote the $t^{\text{th}}$ iterate of this alternating-minimization procedure, $t \geq 0$. At iteration $t+1$, the code matrix $X^{(t+1)}$ is computed based on $H^{(t)}$ through a \textit{sparse coding} step, after which $H^{(t+1)}$ is computed using $X^{(t)}$ through a \textit{dictionary learning} step. %The algorithm alternates between the two steps until a convergence criteria is reached.

For the sparse coding step, Eq.~\ref{eq:l0objective} is combinatorially hard (and nonconvex). %with the objective value having to be assessed for every possible combination of $supp(X)$. 
Instead, several approaches solve an alternate convex objective, with the $\ell_1$ norm replacing the $\ell_0$ quasi-norm. Basis Pursuit (BP) denoising~\cite{Chen2001} and FISTA \cite{beck2009fast} are among the most popular such approaches. More recently, ADMM has been suggested as a more efficient alternative~\cite{cardona2017convolutional}.

Another line of work attempts to solve the original non-convex sparse-coding problem in a greedy fashion using algorithms such as Matching Pursuit (MP)~\cite{Mallat1993} and Orthogonal Matching Pursuit (OMP) \cite{OMP}, a variant of the former. This line of approach is taken throughout this work, as explained next.

\section{Convolutional Dictionary Learning by Convolutional OMP and Convolutional KSVD}
We introduce convolutional OMP (cOMP) for sparse coding, and convolutional KSVD (cKSVD) for convolutional dictionary learning. %While preparing this manuscript we became aware of the unpublished work~\cite{Lecun2010}. 
Our work distinguishes itself from~\cite{Lecun2010} in the use of OMP as opposed to MP in the spare coding step. %This seemingly benign difference has tremendous consequences in the applicability of the theoretical bounds in Section~\ref{sec:theory} on the complexity of spike-sorting in terms of recording length. 
The applicability of recent results from dictionary learning theory~\cite{altmin} rely on results in compressive sensing that have been proved for OMP but not for MP~\cite{tropp2007signal}. %theoretical guarantees  We also demonstrate ways to leverage the convolutional structure to perform memory/computation-efficient matrix operations.

\subsection{An overview of classical OMP and KSVD}
cOMP and cKSVD are used at every iteration of the alternating-minimization procedure, respectively for sparse coding and dictionary learning. Therefore we drop the super-script $t$ indexing the iterates of the procedure and simply refer to $H$ and $X$. Moreover, since the sparse coding step consists of $J$ independent sparse coding problems, we restrict our attention to the case of a single window $Y_j$.
\paragraph{(Sparse Coding)}

OMP is a so-called ``greedy'' algorithm that iteratively selects columns from $H$ to produce an approximation $H\hat{X}_j$ of $Y_j$. Let $t' \geq 1$ be the iteration index of OMP. The algorithm terminates when the approximation error or the sparsity of $\hat{X}_j$ reach a threshold. The inputs of iteration $t'$ are i) the set $S^{(t'-1)}$ of columns that have been selected up to iteration $t'-1$, and ii) the residual error $r^{(t'-1)}$ from projecting $Y_j$ onto the span of $S^{(t'-1)}$.  At iteration 1, $r^{(0)} = Y_j$ and $S^{(0)} = \{\emptyset\}$. Iteration $t'$ of OMP selects the column from $H$ with maximal absolute inner product with $r^{(t'-1)}$. Because the residual is orthogonal to the span of $S^{(t'-1)}$, \emph{a different column of $H$ is selected at every iteration}.

Matching Pursuit is an alternative to OMP that has been used in spike sorting for template matching, specifically to determine the time of action potentials from a putative neuron in extra-cellular recordings~\cite{spyking, kilosort}. MP is different from OMP in that, at iteration $t'$, it computes $r^{(t')} = r^{(t'-1)} - \ip{H_{q^*}}{r^{(t'-1)}}H_{q^*}$, where $q^* = \underset{q \in {1,\cdots,C(W-l+1)}}{\text{argmax}} |\ip{H_{q}}{r^{(t'-1)}}|$. Note the absence of the projection step onto columns that were selected at iterations prior to $t'-1$, which means that the same column can be selected multiple times throughout the algorithm. For spike sorting, this means that MP might detect a spike at the same location more than once. 
%As mentioned previously, theoretical guarantees of sparse recovery for OMP are well studied \cite{tropp2007signal} (including the convolutional case~\cite{Papyan2016}), which justifies the choice of OMP instead of MP.

\paragraph{(Dictionary Learning)}
KSVD \cite{ksvd} is a popular dictionary learning algorithm that updates dictionary elements one at a time. Let $h_k$ be the column of the dictionary $H$ being updated and $x^k$ be the $k^{th}$ row vector of $X$ (It is different from $X^k$ that means $k^{th}$ block of $X$). KSVD uses the SVD to minimize the error between a residual matrix computed from columns other than $k$ and a rank-1 approximation that is the outer product of $h_k$ and $x^k$. More formally, KSVD minimizes%hrough Singular Value Decomposition (SVD), while maintaining the sparsity structure of the coefficients. Specifically, if $d_k$, the column $k$ of a generic dictionary $D$, is being updated, where $x^k$ denotes the row $k$ of $X$,
\begin{equation}
\lVert Y-DX\rVert_\text{F}^2=\Big\lVert Y-\sum_{c=1}^{C}h_cx^c\Big\rVert_F^2=\Big\lVert \underbrace{\Big(Y-\sum_{c\neq k}^{C}h_cx^c\Big)}_{E_k}-h_kx^k\Big\rVert_\text{F}^2
\end{equation}
To maintain the sparsity structure of $x^k$, the columns of $E_k$ corresponding to the support of $x^k$ are extracted to form a shrunk error matrix, $E_k^R$. Finally, SVD is performed on $E_k^R$ to obtain new $\hat{h}_k$ and $\hat{x}^k$. K-SVD cycles through all the dictionary elements in this manner.

\subsection{Sparse coding - Convolutional OMP (cOMP)}
The cOMP involves two computationally intensive steps, namely the inner product step $\langle H^T, r^{(t')}\rangle$, expressed as $H^Tr^{(t')}$, and the least-squares of projecting the residual on the the span of $S^{(t'-1)}$. Considering that $H\in \mathbb{R}^{W\times(\sum_{c=1}^C (W-l+1))}$ is high-dimensional since typical extra-cellular recordings can last on the order of minutes, if not hours, with typical sampling rate of $> 10^4 (Hz)$, the naive projection operation is computationally expensive.

We can take advantage of the convolutional structure of $H$ and compute the projection as a series of $C$ cross-correlation operations as follows \cite{Zhang2008MLSSL},
\begin{equation}
H^T r^{(t')}=\big[(h_1\star r^{(t)})[0], \cdots,(h_1\star r^{(t')})[W-l],\cdots,(h_2\star r^{(t')})[W-l],\cdots,(h_C\star r^{(t')})[W-l]\big]^T
\end{equation}
where $\star$ is a cross-correlation operation. Since this involves only $C$ cross-correlation operations, it is much more efficient than the naive projection. Moreover, there is no need to store the entire matrix $H$ in the memory, as only the filters $\{h_c\}_{c=1}^C$ and $r^{(t')}$ are required.

For the least-squares step, %there is no additional benefit to be gained computationally from the convolutional structure. Instead,
we leverage the fact that $S^{(t'-1)}$ and $S^{(t')}$ are only different by one element and thus the projection onto the span of $S^{(t')}$ can be obtained easily from that onto the span of $S^{(t'-1)}$ to further accelerate OMP computations~\cite{rubinstein2008efficient}. 

%\begin{enumerate}
%	\item CBP (Simoncelli) vs. Convolutional OMP - comparable performance and much easier to implement
%	\item How do we estimate the error threshold for OMP?
%\end{enumerate}
\subsection{KSVD with convolutional dictionary (cKSVD)}
We use the sparse codes $X$ from the cOMP step to perform the convolutional KSVD (cKSVD) step. Motivated by classical KSVD, we will update each $h_c$ at a time. We denote by $\hat{h}_c$ the updated version of $h_c$ following cKSVD and $\widehat{H}_c$ that of $H_c$. Assuming $h_c$ is being updated, % and its convolutional codes, $X^k$ are being updated. Then,
\begin{equation}
\Big\lVert Y-HX\Big\rVert_\text{F}^2=\Big\lVert Y-H_{\backslash c}X^{\backslash c}-H_cX^{c}\Big\rVert_2^2=\Big\lVert E_c - H_cX^c\Big\rVert_2^2
\end{equation}
where $H_{\backslash c}\in \mathbb{R}^{W \times \sum_{j\neq c}(W-l+1)}$ refers to $H$ with $H_c$ removed and $X^{\backslash c}\in\mathbb{R}^{\sum_{j\neq c}(W-l+1)\times J}$ refers to $X$ with $X^c$ removed. The key distinction between cKSVD and classifical KSVD is as follows: since the columns of $H_c$ are the linear shifts of $h_c$, we cannot update each column of $H_c$ independently. Instead, we need to update block-wise to ensure that $\widehat{H}_c$ comprises shifted versions of $\hat{h}_c$. This is done by rearranging $H_{\backslash c}$ and the sparse codes as explained below.

We use $X^c_j\in\mathbb{R}^{W-l+1}$ to denote $x_{c,j}$. Let us assume that $X^c_j$ has $s_{c,j}$ nonzero coefficients, or $s_{c,j}=\big\vert \text{supp}(X_j^c)\big\vert$. The sparsity of $X_j^{c}$ implies that we only need to deal with a subset of $Y_j$, of length $l \times s_{c,j}$, that is \textit{influenced} by $h_c$. Stacking the relevant observations within $Y_j$, and across all $J$ windows, we obtain
\begin{equation}\label{eq:convksvd}
\begin{split}
&\underbrace{\overbrace{\Big[Y_1^{(1)}\Big|\cdots\Big| Y_1^{(s_{c,1})}\Big|\cdots \Big|Y_J^{(s_{c,J})} \Big]}^{\in \mathbb{R}^{l\times \sum_{j=1}^J s_{c,j}}}- \overbrace{\Big[H_{\backslash c}^{(1)}X_{1}^{\backslash c}\Big|\cdots\cdots \Big|H_{\backslash c}^{(s_{c,1})}X_{1}^{\backslash c}\Big|\cdots\Big| H_{\backslash c}^{(s_{c,J})}X_{J}^{\backslash c}\Big]}^{\in \mathbb{R}^{l\times \sum_{j=1}^J s_{c,j}}}}_{E_c}\\
&-h_c\overbrace{\Big[X_1^{(1)}\Big|\cdots\Big|X_1^{(s_{c,1})}\Big|\cdots\Big|X_J^{(s_{c,J})}\Big]}^{\in \mathbb{R}^{\sum_{j=1}^J s_{c,j}}}\\
\end{split}
\end{equation}
where the superscript $(\cdot)$, along with the subscript $j$ for window $j$, denotes the corresponding segment (for $Y_j$) or block (for $H_{\backslash c}$) to nonzero $X_j^{(\cdot)}$. We perform SVD on $E_c$ and assign the first left singular vector as the new dictionary element, $\hat{h}_c$, and easily obtain $\widehat{H}_c$. The new sparse code, $\{\widehat{X}_j^{(s)}\}_{s=1}^{s_{c,j}}$ for $j=1,\cdots,J$, is the first right singular vector multiplied by the first singular value. $\widehat{X}_j$ is constructed by replacing $X_j^{(s)}$ by $\widehat{X}_j^{(s)}$, thereby maintaining the support, $\text{supp}(\widehat{X}_j)=\text{supp}(X_j)$. $H^{(t+1)}$ and updated $X^{(t+1)}$ are obtained after cycling through all $C$ dictionary elements.

\section{How much data does spike sorting require?}
\label{sec:theory}
%Together cOMP and cKSVD form the backbone of a powerful alternating-minimization scheme for convolutional dictionary learning. 
Framing spike sorting as a (convolutional) dictionary learning problem, \emph{solved by alternating minimization}, lets us give theoretical bounds of the amount of data necessary to reliably estimate the waveforms for a given set of neurons. %--and hence the location of the spikes for each waveform by sparse coding following successful dictionary learning. 
In this section, we impose further assumptions on the generative model of Equation~\ref{eq:dtmodel} that allow us to apply results from dictionary learning theory~\cite{altmin}. %to answer the question: how much data does spike sorting require, under reasonable theoretical assumptions?
%With the proposed spike-sorting framework, it is natural to ask what the lower bound for the required amount of data is, in terms of the number of spiking events for each neuron, for the provable guarantees of learning the true waveforms and the locations of spikes. Under some reasonable assumptions, we apply some theorems developed recently in the dictionary learning/sparse coding literature \cite{altmin} to provide the lower bound for the required amount of data.
%-- Demba initial comments
%We put assumptions on waveform templates (incoherence, ie. sufficiently dissimilar), number of neurons allowed to spike simultaneously, that are reasonable in the spike-sorting setting and yield theoretical bounds on how much data is needed for this problem.
We make the following assumptions\\
\noindent \textbf{Assumption 1: Events occur in non-overlapping windows of length $l$}. This assumption is one of mathematical convenience, which allows us to avoid boundary effects.  Mathematically, $\forall c, n_{c,i} = m l$, for some $m$ s.t. $n_{c,i} \leq N-l$.\\
\noindent \textbf{Assumption 2: Refractoriness}. Each neuron has a refractory period of at least $l$ samples ($1$ ms), i.e. $\forall c, n_{c,i}-n_{c,i-1} \neq 0$.
\noindent Under these assumptions, we can express Eq.~\ref{eq:dtmodel} in linear-algebraic form by splitting $y[n]$ into $J = \left \lfloor \frac{N}{l} \right \rfloor$ disjoint windows
\begin{equation}
	Y_j = H X_j, j = 1,\cdots,J
\label{eq:altminmod}
\end{equation}
\noindent where $Y_j \in \R^l$, $X_j \in \R^{C}$, $x_{j,c} \neq 0$ only if neuron $c$ has an event in the $j^{\text{th}}$ window and $H = \begin{bmatrix} h_1 | \cdots | h_c \end{bmatrix} \in \R^{l \times C}$. \\
\noindent \textbf{Assumption 3: At most $s$ neurons can spike simultaneously}. This is equivalent to assuming that $X_j$ from Eq.~\ref{eq:altminmod} is $s$-sparse $\forall j=1,\cdots,J$. If the number of neurons $C > l$, it would be unreasonable to hope to separate up top $C$ neurons. 

Expressing Eq.~\ref{eq:altminmod} in matrix form, i.e. $Y = H X$, let  AltMinDict($Y,H^{(0)},\epsilon_0$) denote the alternating minimization algorithm for dictionary learning from~\cite{altmin}. Under \textbf{Assumptions 1--3}, and the in the absence of noise in Eq.~\ref{eq:dtmodel}, alternating between cOMP and cKSVD reduces to AltMinDict($Y,H^{(0)},\epsilon_0$). We have the following result regarding the complexity of spike sorting
\begin{thm*}
	For each $c=1,\cdots,C$, let $\lambda_c(t)$ denote the conditional intensity function (CIF) of neuron $c$ and $N_c(t)$ the associated counting process. 
Suppose the CIFs are uniformly bounded by $\bar{\lambda}$ Hz. Let $0 < \delta << 1$ be a precision/accuracy parameter. Under assumptions A1--A7 from~\cite{altmin}, with probability at least $1-2\delta$, the $t^{\text{th}}$ iterate $H^{(t)}$ of AltMinDict($Y,H^{(0)},\epsilon_0$) satisfies the following for all $t \geq 1$
\begin{equation}
	\min_{z \in \{-1,1\}} \lVert z h_c^{(t)}-h_c \rVert_2 \leq \frac{1}{2^t} \epsilon_0.
\end{equation}
In particular, Assumption A5 translates into a bound on the recording-length complexity of spike sorting--length of recording required for sorting--of
\begin{equation}
T_r \geq \frac{1}{\bar{\lambda}} 4 \text{max}\left(\frac{C^2}{s},C M^2\right) \text{log}\left(\frac{2C}{\delta}\right) \text{seconds},
\end{equation}
\noindent where $M \approx 1$ is a bound on the normalized spike amplitudes.
	\label{lem:sscmplxt}
\end{thm*}
The theorem states that, as long as the initial dictionary is close to the true one, the error between the iterates of the alternating minimization algorithm for dictionary learning--which convolutional cOMP/cKSVD reduce to with our assumptions above--and the true one will decrease exponential fast.%, as long as the recording length is bigger than a lower bound that depends on (a) the number of neurons $C$, (b) the maximum number of neurons $s$ that are allowed to spike simultaneously and (c) an upper bound $\bar{\lambda}$ on the rate of spiking of the neurons. 
Table~\ref{tab:reclencomp} suggests that the theoretical estimates are very reasonable. %Both our simulations and those from~\cite{altmin} suggest that the term proportional to $C^2$ from the bound is an artifact of the mathematical proofs, so we report the term proportional to $C$. 
The qualitative trend is that the recording length decreases linear with firing rate and increases linearly with the number of neurons. For multiple electrodes that are able to reliably detect the same neurons, the figures should be divided by the number of electrodes. %It is not hard to generalize the figures to the case when the electrode array is so large that subsets of electrodes can be assigned to subsets of neurons~\cite{spyking}. \\
\begin{table}
\caption{Estimates based on theory of the complexity of spike sorting in terms of recording length. We assume $\delta = 0.001$, and at most $s=3$ neurons can spike simultaneously.}
  \centering
    \begin{tabular}{l|*{3}r}
    \toprule
    \diagbox{$\bar{\lambda}$}{$C$} & 5 & 10 & 20 \\
    \midrule
    5 & $45$ secs. & $1$ min. $30$ secs. & $3$ mins. $25$ secs.  \\
    10  & $22$ secs. & $47$ secs. & $1$ min. $42$ secs.  \\
    20 & $11$ secs. & $24$ secs. & $51$ secs. \\
    \bottomrule
    \end{tabular}%
  \label{tab:reclencomp}%
\end{table}%
%\noindent {\color{blue}{ \textbf{Put Table here summarizing result for various values of $C,s,\delta$}}}.

In the \textbf{Supplemental Material}, we re-state assumptions A1--A7 from~\cite{altmin}, discuss their implications and how reasonable they are in the spike-sorting context, and give a sketch of a proof of the theorem.

\section{Application to real and simulated electrode data}
We applied our method to two datasets. We simulated the first dataset using a library of spike waveforms obtained from extra-cellular recordings. The second dataset consists of tetrode recordings from the rat Hippocampus with simultaenous intracellular recording \cite{Henze2000}. For both, we performed $20$ iterations of cOMP \& cKSVD for the dictionary learning step. Following dictionary learning, we used cOMP for spike sorting and standard clustering with K-means as a benchmark. For both data sets, the inputs to the K-means algorithm use  snippets of the signal, that cross a pre-computed threshold,  projected onto the lower dimensional space spanned by the 10 principal components of the snippet matrix with largest singular value. These accounted for $90$\% of the variance~\cite{Lewicki, Simoncelli}. %For both datasets, 10 PCs, which accounted for over 90\% variance, were used for the clustering.

\subsection{Simulated Data: Recovery of the true templates}
We simulated data with three spike waveform templates, each 45 samples long. The data consist of 50 windows, each $3\times 10^4$ samples long (equivalent to $1$ second with $30$ kHz sampling rate). In a single window, the firing rate of each neuron was $40$ Hz and modulated to have peak amplitude distributed uniformly in $[-75, -125]\,(mV)$. We enforced refractoriness by preventing waveforms from the same neurons from overlapping. Finally, we added Gaussian noise with variance corresponding to a desired Signal-to-Noise ratio (SNR).

We initialized the dictionary learning algorithm by perturbing the three templates with additive noise to achieve an average error distance  $err(\hat{h},h)=\max_c\sqrt{1-\langle h_c,\hat{h}_c\rangle^2}$ equal to $0.4$, where $h_c$ and $\hat{h}_c$ as the column $c$ of the two dictionaries of interest, and $\lVert h_c\rVert_2=\lVert \widehat{h}_c\rVert_2=1,\forall c$. 

\begin{figure}
	\centering
	\includegraphics[width=\linewidth]{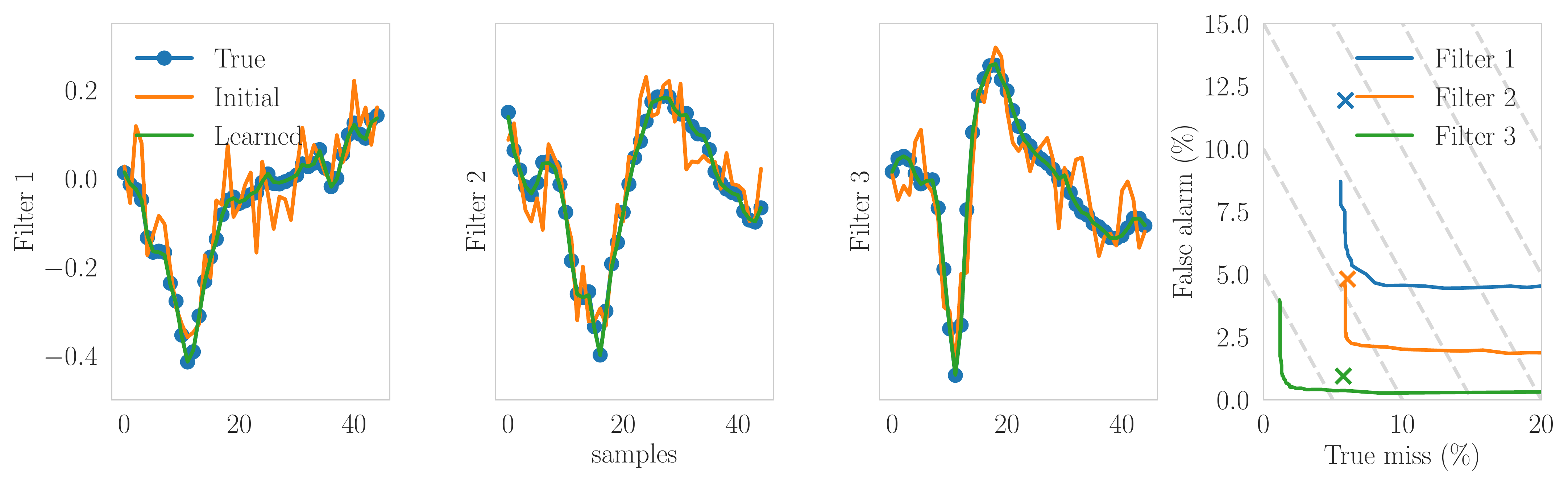}
	\caption{(Left) Dictionary recovery of true templates for the simulated data with SNR 6 dB, with initial template perturbation error distance of 0.4 (Right) False alarm and True miss rate for each filter as the amplitude threshold is varied. The x marker indicates the best K-means 3 cluster result, in terms of the sum of error rates.}
	\label{fig:simulationdatawaveform}
\end{figure}
Fig.\ref{fig:simulationdatawaveform} shows the true/initial/recovered templates after 20 iterations of cOMP \& cK-SVD. The figure show that the algorithm performs exact recovery of the true templates for varying levels of SNR. The learned templates all converged to error distances of less than $10^{-2}$ from the true ones. We also perturbed both the signal and the initial dictionary with higher noise variance and verified that the templates were recovered (results not shown here). 

\subsection{Simulated Data: Recovery of the true sparse codes (spike-sorting)}
Next, we assessed how well cOMP was able to sort spikes from the simulated data at varying levels of SNR. We terminate cOMP when the residual error goes below the noise level that perturbed the signal.

We assessed the performance based on two error statistics: The ``True miss rate" is the proportion of true spikes not identified. The ``False alarm" rate is the proportion of identified spikes that are not true spikes. We assess the rates for the individual templates separately. The error rates were assessed as a function of amplitude threshold, where the crossing of the threshold indicates identification of the recovered waveform as a spike. For noisy data (6 dB), Fig.\ref{fig:simulationdatawaveform} shows that cOMP/cKSVD is able to reduce both error criteria significantly compared to an algorithm that assigns spikes to one of the three clusters found by K-means with $K=3$.% cluster result for the noisy data (6 dB). 
%\vspace{-0.5in}
\subsection{Extra-cellular data from tetrode in rat Hippocampus}
We filtered the data with a highpass filter at $400 Hz$ to remove the slow drift. Additionally, 0.9 second of data with anomalous bursting activities was removed. The cOMP termination criterion was estimated to be the standard deviation of the background noise extracted from a segment that remains below a pre-defined threshold for more than 500 (ms). This automated approach is more appealing than ones in which hyperparameters of the algorithm are manually tuned \cite{spyking, kilosort, Simoncelli}. %\textcolor{blue}{Mention that based on PC might not be the same Harris dataset as others?}
 
%Estimation of this hyperparameter directly from the data is more plausible and automated than manually having to fine-tune the hyperparameters as in \cite{kilosort, Simoncelli}, that does not directly involve the background noise.

\begin{figure}
	\centering
	\includegraphics[width=0.85\linewidth]{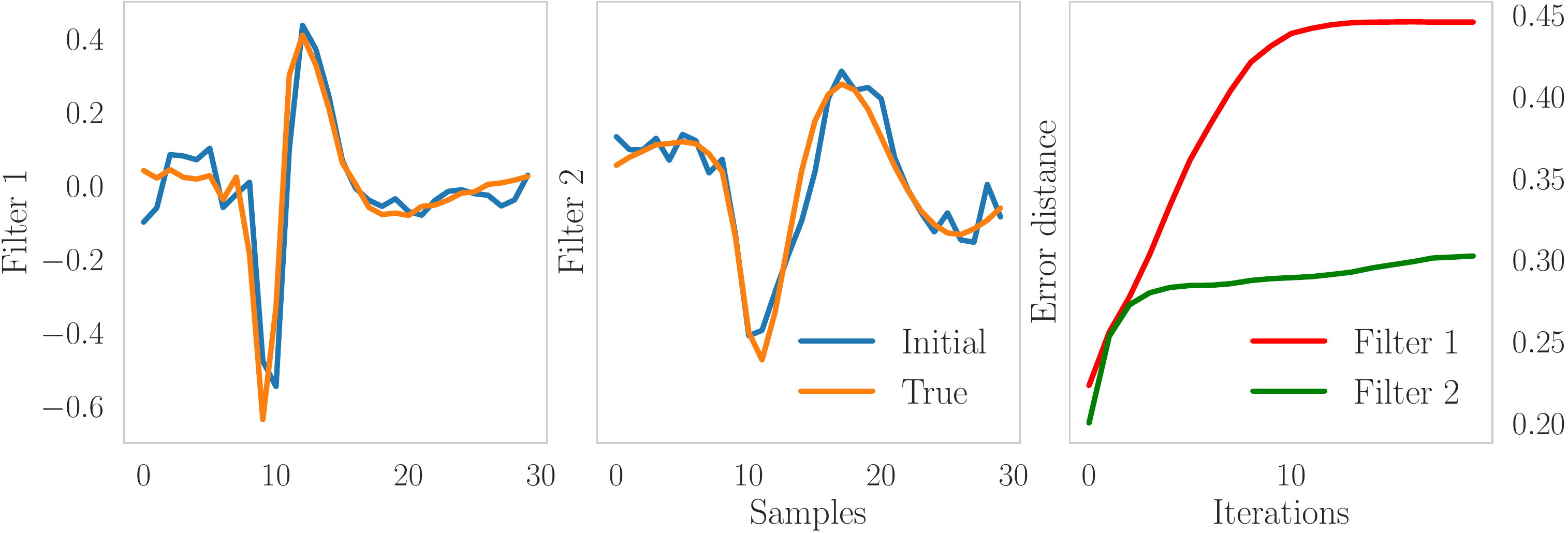}
	\caption{(Left) Initial and learned dictionary for the real data for $C=2$. (Right) The error distance betwen the initial and the learned dictionary as a function of iterations.}
	\label{fig:realdatawaveform}
\end{figure}

\begin{figure}
	\centering
	\includegraphics[width=\linewidth]{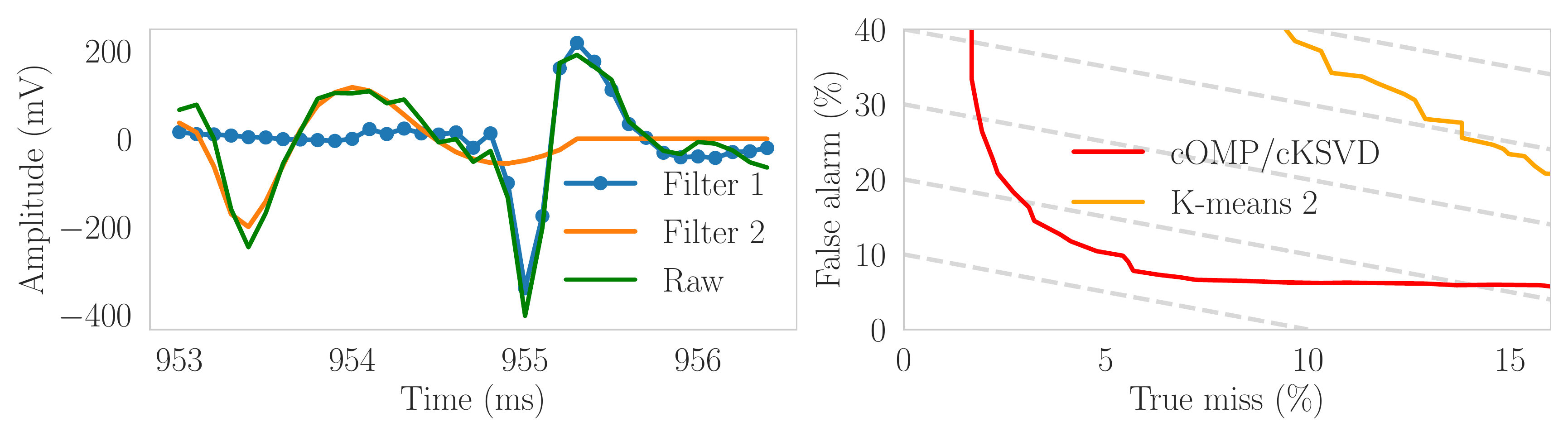}
	\caption{(Left) Example of how cOMP decomposes the raw signal into sum of two learned waveforms. (Right) The false alarm and true miss rate for cOMP/cKSVD and K-means with 2 clusters with varying threshold.}
	\label{fig:realdataperformance}
\end{figure}

We ran cOMP/cKSVD for $C=2$. For the initialization of the templates (30 samples), we randomly picked the $C$ segments of the signal that crossed the threshold that were as distant from each other as possible, in terms of the error distance defined previously. 
%(Figures for $C=3,5$ in the \textbf{Supplemental Material}.)
Fig.\ref{fig:realdatawaveform} and Fig.\ref{fig:realdataperformance} show the result for $C=2$. The error distance displayed is that between the initial and the learned dictionary, and is expected to increase as the dictionary is learned. The fact that the error stabilizes after a certain number of iterations indicates that it has learned a set of templates that is deemed optimal. The true miss of $ <2\,\%$ (in Fig.\ref{fig:realdataperformance}) in cOMP/cKSVD, which is a significant improvement over the K-means algorithm, is comparable to existing methods. Finally, Fig.\ref{fig:realdataperformance} is an example demonstrating that a shifted linear combination of templates is able to reconstruct the raw signal accurately.

\section{Conclusion}
We have cast the spike-sorting problem in the framework of convolutional dictionary learning and showed that it can be solved efficiently through an iterative procedure that alternates between convolutional OMP and convolutional KSVD, generalizations respectively of OMP and KSVD. The framing of spike sorting as a convolutional dictionary learning problem, and its solution via alternating minimization, let us employ recent theoretical results in the field of dictionary learning that give estimates on the length of recordings required for spike sorting. %Convolutional OMP is  simple algorithm that exploits the convolutional structure of the problem for computational efficiency. 
In future work, we will 1) massively parallelize the sparse coding step with GPU architecture, 2) generalize our framework to two dimensions to process multi-electrode recordings where the spatial extent of the areas must be considered and 3) derive improved theoretical bounds for convolutional dictionary learning.

%\noindent \textcolor{blue}{A simple remedy for efficiency is parallelization of OMP across disjoint windows of the data. Extending this further, we have implemented the Fast Iterative Soft-Thresholding Algorithm (FISTA) that can be run on GPU.}

%\begin{enumerate}
%	\item Need for better initialization schemes. Spectral clustering
%\end{enumerate}

\newpage

\bibliographystyle{unsrt}
%\bibliography{spikesorting}
\bibliography{nips18}

\newpage

\section{Supplemental Material}

\subsection{Assumptions A1--A7 from~\cite{altmin}}

\begin{itemize}
	\item[(A1)] \textbf{Dictionary Matrix satisfying RIP}: The dictionary matrix $H$ has $2s$-RIP constant of $\delta_{2s} < 0.1$.
	\item[(A2)] \textbf{Spectral Condition of Dictionary Elements:} The dictionary matrix has bounded spectral norm, for some constant $\mu_1 > 0$, $\lVert H \rVert_2 < \mu_1 \frac{C}{l}$.
	\item[(A3)] \textbf{Non-zero Entries in Coefficient Matrix}: The non-zero entries of $X$ are drawn i.i.d. from a distribution such that $E[(X_j^c)^2]$ = 1, and satisfy the following a.s.: $|X_j^c| \leq M, \forall, c,j$.
	\item[(A4)] \textbf{Sparse Coefficient Matrix}: The columns of the coefficient matrix have $s$ non-zero entries which are selected uniformly at random from the set of all $s$-sized subsets of $\{1,\cdots,C\}$. It is required that $s \leq \frac{l^{1/6}}{c_2 \mu_1^{1/3}}$, for some universal constant $c_2$.
	\item[(A5)] \textbf{Sample Complexity}: For some universal constant $c_3 = 4$, and a given failure parameter $\delta > 0$, the number of samples $J$ needs to satisfy
	\begin{equation}
	J \geq c_3 \text{max}(C^2,C M^2 s)\text{log}\left(\frac{2C}{\delta}\right).
	\end{equation}
	\item[(A6)] \textbf{Initial dictionary with guaranteed error bound}: It is assumed that we have access to an initial dictionary estimate $H^{(0)}$ such that 
	\begin{equation}
	\max_{c \in \{1,\cdots,C\}}\min_{z \in \{-1,1\}} \lVert z h_c^{(0)}-h_c \rVert_2 \leq \frac{1}{2592s^2}.
	\end{equation}
	\item[(A7)] \textbf{Choice of Parameters for Alternating Minimization}:  AltMinDict($Y,H^{(0)},\epsilon_0$) uses a sequence of accuracy parameters $\epsilon_0 = \frac{1}{2592s^2}$ and
	\begin{equation}
	\epsilon_{t+1} = \frac{25050\mu_1 s^3}{\sqrt{l}}\epsilon_t.
	\end{equation}
\end{itemize}

\noindent \textbf{Interpretation of A1--A7 for spike sorting}. We now discuss the appropriateness of these assumptions in the spike-sorting setting. As with most theoretical results, the theorem provides a set of guidelines that appear reasonable (Table 1)

\begin{itemize}
	\item[1.] A1 requires the RIP constant of order $2s$ for $H$ to be smaller than a small value. Loosely, this states that subset of colums of $H$ of size $2s$, i.e. subsets of the $C$ neural templates of order $2s$ should be dissimilar. This is an assumption that is hard to satisify in spike sorting because the neural templates can be very similar. The authors in~\cite{altmin} relax the RIP assumption to incoherence, i.e. an upper bound of the inner-product between pairs of neural templates. The larger this upper bound, the smaller then number of neurons that can spike simultaneously while guaranteeing exact recovery. The bound for the incoherent case is of the same form as that above. Suppose  $ M_\ell \leq |X_j^c| \leq M_u = M$, the main difference is that one has to pay a factor $\left(\frac{M_u}{M_\ell}\right)^2$ as opposed to $M^2$, i.e. a factor proportional to the ration of the maximum to the minimum normalized spike amplitudes. Assuming the spike amplitude distribution is fairly concentrated around some mean (high SNR), this is negligible. 
	\item[2.] A2 is a reasonable assumption on the largest eigenvalue of $H$ from Eq. 9.
	\item[3.] A3 can be generalized to requiring that the amplitudes of the coefficients $E[(X_j^c)^2] = \sigma_x^2$ i.e. have bounded variance, and an upper bound on the absolute value of the coefficients normalized by their standard deviation, i.e. $\frac{|X_j^c|}{\sigma_x^2} \leq M$, assumptions which are also reasonable in the spike-sorting setting.
	\item[4.] A4 requires that only a few neurons are allowed spike simultaneously. Loosely, it states that $s^3 \propto l^{1/2}$.
	\item[5.] A5 gives the sample complexity that guarantees that, with the stated probability, the bound from the theorem holds. The form of the bound comes from the concentration results from random matrix theory used in~\cite{altmin} to prove the result that alternating minimization succeeds with high probability for dictionary learning. %stating that A3 and A4 yield bounds for the probability of the event in the Theorem in terms of $n, M, c, S$. Fixing $\delta$ allows to solve for $n$ and gives A5.
	\item[6.] A6 states that the initial dictionary should be close to the true one, an assumption which is reasonable in spike sorting.
	\item[7.] We can re-write the matrix form of Eq. 9 as follows $Y = HX = H^{(t)} X + (H-H^{(t)})X$. We can treat the second term, that comes from approximating $H$ with $H^{(t)}$, as noise. A7 states that, as the iterations of cOMP and cKSVD proceed, we should decrease the noise level in OMP, which is reasonable. In practice, this did not affect results much in the high SNR case.
	
\end{itemize}

\subsection{Proof of Theorem}

We give a sketch of a proof of the Theorem stated in the main manuscript.

\begin{proof}
	The theorem follows from applying Theorem 1 from~\cite{altmin}. Under assumptions A1--A7, this gives a bound for $J \geq 4 \text{max}(C^2,C M^2 s)\text{log}\left(\frac{2C}{\delta}\right)$, i.e. the sample complexity of spike-sorting cast as dictionary learning (under our assumptions above). $J$ corresponds to the number of windows of length $l$ for which at least one of the neurons spikes.  We can turn this into a recording-length complexity by upper bounding the rate of occurrence of the event ``at least one of the neurons spikes''. Let $\lambda(t)$ denote the rate of occurrence of said event. 
	
	In an interval of width $\Delta = 1$ ms, we can use a union bound argument to upper bound the probability of the event $V = \{\cup_{c=1}^C V_c\}$, $V_c = \{N_c(t+\Delta)-N_c(t) = 1\}$ in terms of the CIFs of the neurons and $\Delta$
	\begin{equation}
	P[V] \leq \sum_{c=1}^C P[V_c] = \sum_{c=1}^C \lambda_{c}(t) \Delta \leq s \bar{\lambda}\Delta,
	\end{equation}
	where $s$ comes from the fact that we alllow at most $s$ neurons to spike at the same time. Therefore, by definition of the CIF, $\lambda(t) = \lim_{\Delta \to 0} \frac{P[V]}{\Delta} \leq s \bar{\lambda}$, yielding the recording-length complexity $\frac{J}{s\bar{\lambda}}$, as stated in the theorem.
\end{proof}

\end{document}